\documentclass[runningheads]{llncs}
\usepackage{graphicx}
\usepackage{amsmath}
\usepackage{amsfonts}
\usepackage{multirow}
\usepackage{color} 

\usepackage{tikz}
\usetikzlibrary{calc}
\usetikzlibrary{arrows.meta}
\usetikzlibrary{shapes.geometric}
\begin{document}
\title{Adversarial Domain Feature Adaptation for Bronchoscopic Depth Estimation}
\titlerunning{Adversarial Domain Feature Adaptation for Bronchoscopic Depth}
\author{
	Mert Asim Karaoglu\inst{1, 2} \and
	Nikolas Brasch\inst{2} \and
	Marijn Stollenga\inst{1} \and
	Wolfgang Wein\inst{1} \and
	Nassir Navab\inst{2, 3} \and
	Federico Tombari\inst{2, 4} \and
	Alexander Ladikos\inst{1}
}

\authorrunning{M.A. Karaoglu et al.}
\tocauthor{M.A. Karaoglu}
\institute{
	$^{1}$ {ImFusion GmbH, Munich, Germany}\\
	$^{2}$ {Computer Aided Medical Procedures, Technische Universität München, Munich, Germany}\\
	$^{3}$ {Computer Aided Medical Procedures, Johns Hopkins University, Baltimore, Maryland USA}\\
	$^{4}$ {Google, Zurich, Switzerland}\\
}
%
%
\maketitle              
\begin{abstract}
Depth estimation from monocular images is an important task in localization and 3D reconstruction pipelines for bronchoscopic navigation.
Various supervised and self-supervised deep learning-based approaches have proven themselves on this task for natural images.
However, the lack of labeled data and the bronchial tissue's feature-scarce texture make the utilization of these methods ineffective on bronchoscopic scenes.
In this work, we propose an alternative domain-adaptive approach.
Our novel two-step structure first trains a depth estimation network with labeled synthetic images in a supervised manner; then adopts an unsupervised adversarial domain feature adaptation scheme to improve the performance on real images.
The results of our experiments show that the proposed method improves the network's performance on real images by a considerable margin and can be employed in 3D reconstruction pipelines.

\keywords{Bronchoscopy \and Depth estimation  \and Domain adaptation.}
\end{abstract}

\section{Introduction}
Lung cancer is the leading cause of all cancer-related deaths, accounting for 24\% of them in the US in 2017 \cite{siegel2020cancer}.
The statistics highlight that the patient's survival profoundly depends on the disease stage, intensifying the importance of early diagnosis \cite{siegel2020cancer}.
Diminishing the potential risks of more invasive techniques, transbronchial needle aspiration (TBNA) is a modern approach for pulmonary specimen retrieval for a decisive diagnosis \cite{liu2015evolution}. 
Conducted by operating a bronchoscope to reach the suspected lesion site, segmented on a pre-operative CT scan, navigation is an existing challenge for TBNA procedures since it requires registration to the pre-operative CT plan.

One way to achieve this is to register electromagnetic (EM) tracking data captured from the bronchoscope to the segmented airway tree of the pre-operative CT-scan \cite{schwarz2006real,hofstad2017intraoperative,lavasani2021bronchoscope}.
In addition to the sensory errors caused by electromagnetic distortion, anatomical deformations are a principal challenge for EM-based approaches.

Motivated by the higher resilience of the camera data against such deformations, researchers have focused on vision-based approaches in recent years. 
Inspired by their success in natural scenes, direct and feature-based video-CT registration techniques \cite{sinha2018endoscopic} and simultaneous localization and mapping (SLAM) pipelines \cite{visentini2017deep,chen2018slam} have been investigated in various studies. Even though these approaches have shown a certain level of success, the feature-scarce texture and the photometric-inconsistencies caused by specular reflections are found to be common challenges.

\definecolor{Fs}{HTML}{00a1d4}
\definecolor{Fr}{HTML}{846dca}
\definecolor{Gs}{HTML}{cc5156}
\definecolor{A}{HTML}{4ea77e}

\tikzset{Fs/.style={black,draw=black,fill=Fs,regular polygon, regular polygon sides=3,minimum height=0.1cm}}
\tikzset{Fr/.style={black,draw=black,fill=Fr,regular polygon, regular polygon sides=3,minimum height=0.1cm}}
\tikzset{Gs/.style={black,draw=black,fill=Gs,regular polygon, regular polygon sides=3,minimum height=0.1cm}}
\tikzset{A/.style={black,draw=black,fill=A,regular polygon, regular polygon sides=3,minimum height=0.1cm}}
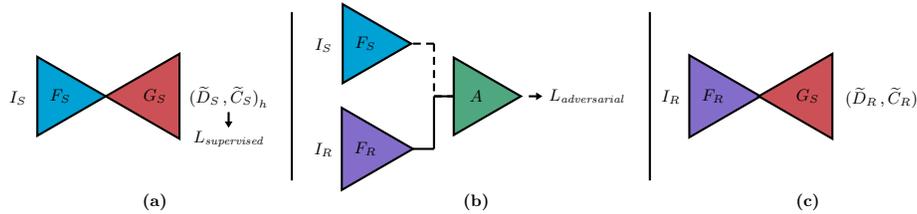
\begin{figure}[t]
	\centering
		\begin{tikzpicture}	[thick, scale=0.7, every node/.style={scale=0.7}]
		
		
		\node[Fs, label={[xshift=-0.8cm, yshift=-0.8cm] $I_S$}, shape border rotate=-90,minimum width=1.7cm, minimum height=0.1cm] (AEncoderS) at (0,0) {$F_{S}$};
		\node[Gs, label={[xshift=1.4cm, yshift=-0.9cm] ${(\widetilde{D}_S \, \mathpunct{,} \widetilde{C}_S)}_h$}, shape border rotate=90,minimum width=1.7cm, minimum height=0.1cm] (ADecoderS) at (1.8,0) {$G_{S}$};
		
		\node[label] (ALSupervised) at (3.2, -0.8) {$L_{\textit{supervised}}$};
		
		\draw[black, arrows={-Triangle[scale=0.5]}]  (3.2, -0.3) -- (3.2, -0.6);
		
		
		\draw[black, arrows={-[scale=0.5]}]  (4.4, 1.6) -- (4.4, -1.6);
		
		
		\node[Fs, label={[xshift=-0.8cm, yshift=-0.8cm] $I_S$}, shape border rotate=-90,minimum width=1.7cm, minimum height=0.1cm] (BEncoderS) at (5.8, 1) {$F_{S}$};
		
		\node[Fr, label={[xshift=-0.8cm, yshift=-0.8cm] $I_R$}, shape border rotate=-90,minimum width=1.7cm, minimum height=0.1cm] (BEncoderR) at (5.8, -1) {$F_{R}$};
		
		\coordinate (CBEncoderS0) at ($ (BEncoderS) - (-0.9, 0) $);
		\coordinate (CBEncoderS1) at ($ (CBEncoderS0) - (-0.4, 0) $);
		\draw[densely dashed, black]  (CBEncoderS0) -- (CBEncoderS1);
		\coordinate (CBEncoderS2) at ($ (CBEncoderS1) - (0, 1.0) $);
		\draw[densely dashed, black]  (CBEncoderS1) -- (CBEncoderS2);
		\coordinate (CBEncoderS3) at ($ (CBEncoderS2) - (-0.4, 0) $);
		\draw[densely dashed, black]  (CBEncoderS2) -- (CBEncoderS3);
		
		\coordinate (CBEncoderR0) at ($ (BEncoderR) - (-0.9, 0) $);
		\coordinate (CBEncoderR1) at ($ (CBEncoderR0) - (-0.4, 0) $);
		\draw[black]  (CBEncoderR0) -- (CBEncoderR1);
		\coordinate (CBEncoderR2) at ($ (CBEncoderR1) - (0, -1.0) $);
		\draw[black]  (CBEncoderR1) -- (CBEncoderR2);
		\coordinate (CBEncoderR3) at ($ (CBEncoderR2) - (-0.4, 0) $);
		\draw[black]  (CBEncoderR2) -- (CBEncoderR3);
		
		\coordinate (CBDiscriminator) at ($ (CBEncoderR3) - (-0.4, 0) $);
		\node[A, shape border rotate=-90,minimum width=1.7cm, minimum height=0.1cm] (BDiscriminator) at (CBDiscriminator) {$A$};
		
		\coordinate (CBDiscriminator0) at ($ (CBDiscriminator) - (-1, 0) $);
		\node[label] (BLAdversarial) at (10.0, 0.0) {$L_{\textit{adversarial}}$};
		\draw[black, arrows={-Triangle[scale=0.5]}]  (CBDiscriminator0) -- (BLAdversarial);
		
		
		\draw[black, arrows={-[scale=0.5]}]  (11.2, 1.6) -- (11.2, -1.6);
		
		\node[Fr, label={[xshift=-0.8cm, yshift=-0.8cm] $I_R$}, shape border rotate=-90,minimum width=1.7cm, minimum height=0.1cm] (CEncoderR) at (12.4,0) {$F_{R}$};
		\node[Gs, label={[xshift=1.4cm, yshift=-0.9cm] ${(\widetilde{D}_R \, \mathpunct{,} \widetilde{C}_R)}$}, shape border rotate=90,minimum width=1.7cm, minimum height=0.1cm] (CDecoderS) at (14.2,0) {$G_{S}$};
		
		\coordinate (CA) at ($ (ADecoderS) - (0, 2) $);
		\node[label] (A) at (CA) {$\textbf{(a)}$};
		
		\coordinate (CB) at ($ (BDiscriminator) - (0, 2) $);
		\node[label] (B) at (CB) {$\textbf{(b)}$};
		
		\coordinate (CC) at ($ (CDecoderS) - (0, 2) $);
		\node[label] (C) at (CC) {$\textbf{(c)}$};

	\end{tikzpicture}
	\caption{Outline of the proposed domain-adaptive pipeline. (a) Supervised training of the encoder-decoder structure using the synthetic images ($ I_S $), predicted depth images ($ \widetilde{D}_S $) and confidence maps ($ \widetilde{C}_S $). (b) Adversarial training scheme to train a new encoder ($ F_R $) for the images from the real domain ($ I_R $). During the optimization, the weights are updated only on the flow drawn with the solid line. (c) For inference on the real domain, $ F_R $ is connected to the decoder trained in the first step ($ G_S $).}
	\label{fig:method_network}
\end{figure}

The shortcomings of the direct and feature-based methods have led researchers to focus on adopting depth information to exploit the direct relationship with the scene geometry. 
Following the advancements in learning-based techniques, supervised learning has become a well-proven method for monocular depth estimation applied to natural scenes. 
However, it is challenging to employ for endoscopy tasks due to the difficulty of obtaining ground-truth data. An alternative way is to train the network on synthetic images with their rendered depth ground-truths. But due to the domain gap between the real and synthetic images, these models tend to suffer from a performance drop at inference time.
To address this \cite{chen2019slam} utilizes Siemens VRT \cite{eid2017cinematic} to render realistic-looking synthetic data for supervised training. However, the renderer is not publicly available and it has only been demonstrated on colonoscopy data. \cite{mahmood2018unsupervised} employs a separate network for an image-level transfer before utilizing the depth-estimation model.
Aside from the second network's additional runtime cost, executed at the more complex image-level it is not easy to ensure the preservation of the task-related features during the domain transfer.
Adopting a fully unsupervised approach, \cite{Shen2019} employs a CycleGAN \cite{zhu2017unpaired} based method for bronchoscopic depth estimation from unpaired images. However, CycleGANs focus on the image appearance and might not preserve the task-relevant features. In \cite{liu2019dense,liu2020reconstructing} a structure-from-motion approach is used to obtain a sparse set of points to supervise a dense depth estimation network for sinus endoscopy scenes. While this works well for nasal and sinus passages with their feature rich texture, it is much more challenging to apply to feature-scarce bronchoscopic data.

In this work, we therefore propose a novel, two-step domain-adaptive approach, for monocular depth estimation of bronchoscopic scenes, as outlined in Figure \ref{fig:method_network}.
Overcoming the lack of labeled data from the real domain, the first step trains a network utilizing synthetic images with perfect ground-truth depths.
In the second step, inspired by \cite{vankadari2020unsupervised}, we employ an unsupervised adversarial training scheme to accommodate the network for the real images, tackling the domain adaptation problem at the task-specific feature-level.
Our method requires neither the ground-truth depth nor synthetic pairs for the real monocular bronchoscopic images for training.
Moreover, unlike some of the explicit domain transfer methods, it does not need a secondary network at inference, improving the overall runtime efficiency.
To evaluate our method, we develop a CycleGAN-based explicit domain transfer pipeline connected to the supervised network trained only on synthetic data in the first step.
We conduct a quantitative test on a phantom and a qualitative test on human data. Furthermore, we employ our method in a 3D reconstruction framework to assess its usefulness for SLAM-based navigation.

\section{Method}
\subsection{Supervised Depth Image and Confidence Map Estimation}
Targeting optimal information compression at their code, encoder-decoder models, such as U-net \cite{ronneberger2015u}, are a favored choice for various pixel-wise regression tasks.
Aiming at an optimal point between accuracy and runtime performance, we utilize a U-net variant \cite{godard2019digging} with a ResNet-18 \cite{he2016deep} encoder.
On the decoder side, a series of nearest neighbor upsampling and convolutional layers are configured to regain the input's original size. 
After each upsampling operation, the corresponding features from the encoder level are concatenated to complete the skip-connection structure. 

Motivated by \cite{facil2019cam}, we alter this architecture with the addition of coordinate convolution layers \cite{liu2018intriguing}. 
Coordinate convolution layers introduce an additional signal that describes a connection between the spatial extent and the values of the extracted features, resulting in a faster convergence for supervised training \cite{liu2018intriguing}. 
In total, we employ five coordinate convolution layers at the skip connections and the bottleneck, right before connecting to the decoder.

The decoder's last four levels' outputs form the scaled versions of the estimated depth images and the confidence maps. 
We utilize the output set in a multi-scale loss consisting of three components: depth estimation loss, scale-invariant gradient loss, and confidence loss.

For depth regression, our objective is to estimate the depth values in the original scale of the ground-truth ($ D $).
As the pixel-wise error, we employ the BerHu loss ($ \mathcal{B} $) as defined in \cite{laina2016deeper}:
\begin{equation}
	L_{\textit{depth}}(D, \widetilde{D})=\sum_{i, j}\mathcal{B}(\mid D(i, j)-\widetilde{D}(i, j)\mid, c)
	\label{eq:loss_depth}
\end{equation} 
where $ D(i, j) $  and $  \widetilde{D}(i, j) $  are the ground-truth and the predicted depth values at the pixel index $ (i, j) $. 
The threshold $ c $ is computed over a batch as:
\begin{equation}
	c=k \max _{t, i, j}\left(|D^t(i, j)-\widetilde{D}^t(i, j)|\right)
	\label{eq:threshold_berhu}
\end{equation} 
where $ k $ is set to $ 0.2 $ as in \cite{laina2016deeper} and $ t $ is an instance of the depth images inside the given batch.

To ensure smooth output depth images, we employ the scale-invariant gradient loss as introduced in \cite{ummenhofer2017demon} as:
\begin{equation}
	L_{\textit{gradient}}(D, \widetilde{D}, h)=\sum_{i, j}\left\|g(D(i, j), h)-g(\widetilde{D}(i, j), h)\right\|_{2}
	\label{eq:loss_gradient}
\end{equation} 
where $ g $ is the discrete scale-invariant finite differences operator \cite{ummenhofer2017demon} and $ h $ is the step size.

Inspired by \cite{facil2019cam}, we use a supervised confidence loss. 
To provide the supervision signal, the ground-truth confidence map is calculated as:
\begin{equation}
	C(i, j)=e^{-|D(i, j)-\widetilde{D}(i, j)|}
	\label{eq:confidence_ground_truth}
\end{equation} 
Based on this, we define the confidence loss to be the $ \mathcal{L}_1 $ norm between the ground truth and the prediction as:
\begin{equation}
	L_{\textit{confidence}}(C, \widetilde{C})=\sum_{i, j}|C(i, j)-\widetilde{C}(i, j)|
	\label{eq:loss_confidence}
\end{equation} 

To form the total loss, we combine the three factors with a span over the four different output scales as:
\begin{equation}
	\begin{split}
		L_{supervised}(D, \widetilde{D}_h)=\sum_{h \in{\{1, 2, 4, 8\}}} (& \lambda_{\textit{depth}}L_{\textit{depth}}(D, u_h(\widetilde{D}_h))\\
		&+  \lambda_{\textit{gradient}}L_{\textit{gradient}}(D, u_h(\widetilde{D}_h), h)\\
		&+  \lambda_{\textit{confidence}}L_{\textit{confidence}}(C, u_h(\widetilde{C}_h)))
		\label{eq:supervised}
	\end{split}
\end{equation} 
$ \lambda $s are the hyper-parameters to weight each factor, $ h $ is the the size ratio of the ground-truth to the predicted depth images, and $ u_h $ is the bilinear upsampling operator that upsamples an image by a scale of $ h $.

\subsection{Unsupervised Adversarial Domain Feature Adaptation}

\paragraph{}

In the second step of the training pipeline, we incorporate the pre-trained encoder $ F_S $ to adversarially train a new encoder $ F_R $ for the images from the real domain.
For this task, we empirically decided to utilize three separate PatchGAN discriminators \cite{isola2017image} ($ A^i $, where $ i\in\{1, 2, 3\} $) to be employed at the encoder's last two skip-connections and the bottleneck, after the coordinate convolution layers, to reduce the domain gap at the task-specific feature levels. 
At inference, the new encoder, $ F_R $, is connected to the previously trained decoder $ G_S $ for depth image and confidence map estimation.
Like other neural network models, GANs have limited learning capacity \cite{goodfellow2014generative}. 
Trained with a lack of direct supervision, it is oftentimes inevitable for GANs to fall into local minima that are not an optimal hypothesis, referred to as a mode collapse.
To increase the model's resilience against this phenomena, we initialize the new encoder $ F_R $ with the same weights as the pre-trained encoder $ F_S $.
Furthermore, in addition to its improvement for convergence time in supervised learning, we use coordinate convolution layers to lower the chance of a possible mode collapse in adversarial training \cite{liu2018intriguing}.

For each discriminator, we employ the original GAN loss \cite{goodfellow2014generative}.
The total adversarial loss is the sum of  discriminator and encoder losses across all $ i $, where $ i \in\{1, 2, 3\} $ is the index of the encoder's feature level ($ F^i $) and its corresponding discriminator ($ A^i $):
\begin{equation}
	\begin{split}
		L_{\textit{adversarial}}(A, F_S, F_R, I_S, I_R)=\sum_{i \in{\{1, 2, 3\}}}(&L_{\textit{discriminator}}(A^i, F_S^i, F_R^i, I_S, I_R) \\
		&+ L_{\textit{encoder}}(A^i, F_R^i, I_R))
		\label{eq:loss_adversarial}
	\end{split}
\end{equation}

\section{Experiments}
\subsubsection{Data.}
\begin{figure}[t]
	\centering
	\includegraphics[width=1\linewidth]{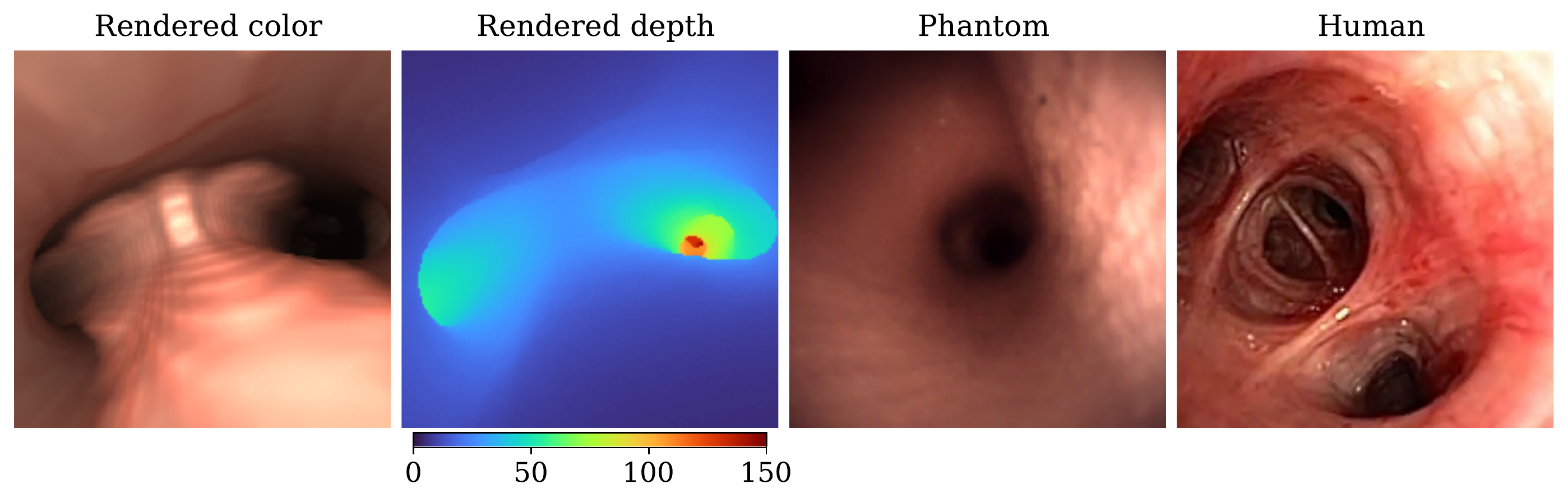}
	\caption{Sample images from our renderings, phantom recordings, and human dataset acquired from \cite{youtube}. The depth values are in \textit{mm}.}
	\label{fig:sample}
\end{figure}
For our experiments, we use three different datasets: synthetic renderings, recordings inside a pulmonary phantom, and human recordings.
The synthetic dataset consists of 43,758 rendered color ($ I_S $) and depth ($ D_S $) image pairs.
For the supervised training, we randomly split twenty percent of it to be utilized as the validation set.
The airway tree used for data generation is segmented from the CT scan of a static pulmonary phantom and rendered using a ray-casting volume renderer.
We model the virtual camera after our bronchoscope's intrinsic properties.

The training split of the pulmonary phantom dataset consists of 7 sequences, in total 12,720 undistorted frames, recorded inside the same phantom.
For the quantitative evaluation, we use a separate set with known 3D tracking information acquired with the bronchoscope's EM sensor.
Registered to the phantom's segmented airway tree, we manually verify and pick 62 correctly registered frames and render their synthetic depth images as the ground-truth.

Our human recordings consist of two sets. 
For training, we use our in-house dataset of three sequences with a total of 16,502 frames. 
For qualitative testing, a set of 240 frames are acquired from a publicly available video \cite{youtube}.
All of our input color images and outputs are of size $ 256 $ by $ 256 $.

\subsubsection{Experiment Setup.}
The networks are implemented on PyTorch 1.5 \cite{paszke2017automatic}.
For the supervised training, we employ random vertical and horizontal flips and color jitters as the data augmentations.
For the adversarial training, we only apply the random flips to the real domain but keep the rest for the synthetic images.
In the first step, we train the supervised model for 30 epochs using the synthetic dataset.
At the output, we employ a ReLU activation \cite{nair2010rectified} for the depth image, as in \cite{laina2016deeper}, and a sigmoid activation for the confidence map.
We set the batch size to 64 and utilize the Adam optimizer \cite{kingma2014adam} with a learning rate of $ 10^{-3} $. 
In the second step, the new encoder is domain-adapted to the real images with a training of 12,000 iterations.
For the adversarial training, we use the Adam optimizer with the learning rate set to $ 5(10)^{-6} $, and it is halved at three-fifth and four-fifth of the total iterations.
In both cases, we set the $ \beta_1 $ and $ \beta_2 $ values of the Adam optimizers to their defaults, $ 0.9 $ and $ 0.999 $.

To have a baseline comparison against an explicit domain transfer approach, we train two standard CycleGANs \cite{zhu2017unpaired} (one for the phantom and the other for the human data) and transfer the real images to the synthetic domain. 
We use these outputs as the input to our model trained only on synthetic data to complete the pipeline for depth and confidence estimation.

We employ our model in a 3D reconstruction framework based on \cite{park2017colored,choi2015robust}. 
The method essentially employs multiple pose graphs to approach the problem from local and global perspectives using feature-based tracking information and colored ICP \cite{park2017colored} for a multi-scale point cloud alignment. 
Unlike \cite{park2017colored,choi2015robust}, we do not consider loop closures in our settings because our test sequences are small in size and do not contain loops.

\subsubsection{Quantitative Results.}
In Table \ref{tab:results_quantitative}, we evaluate the proposed method against the vanilla model (only trained on synthetic images in a supervised manner) and the explicit domain transfer approach. 
Moreover, in this test, we compare these models against their versions without coordinate convolution layers.
The results reveal that our proposed method performs better across all metrics than the others.
Additionally, we confirm that the use of the coordinate convolution layers significantly improves the adversarial training step of our approach.
\begin{table}[b]
	\caption{Quantitative analysis on our pulmonary phantom dataset. The depth values are estimated in \textit{mm}. The ground-truth depth values for the test are in the range of $ 1.74 \text{mm} $ and $ 142.37 \text{mm} $. The best value for each metric is shown in \textbf{bold} characters.}
	\label{tab:results_quantitative}
	\begin{center}
		\resizebox{0.95\textwidth}{!}{%
		\begin{tabular}{ c c c c c c }
			\hline \multirow{2}{*} {\textbf { Model }} & \multirow{2}{*} {\textbf { Mean abs. rel. diff. }} & \multirow{2}{*} {\textbf { RMSE }} & \multicolumn{3}{c} {\textbf { Accuracy }} \\
			\cline { 4 - 6 } & & & $ \sigma=1.25 $ & $ \sigma=1.25^{2} $ & $ \sigma=1.25^{3} $\\
			\hline
			\multicolumn{6}{c} {\textit { Without coordinate convolution }} \\
			\hline
			Vanilla &  $ 0.659 $ & $ 8.407 $ & $ 0.315 $ & $ 0.570 $ & $ 0.733 $\\
			Explicit DT & $ 0.524 $ & $ 8.731 $ & $ 0.293 $ & $ 0.619 $ & $ 0.810 $\\
			Domain-adapted &$ 0.543 $ & $ 8.505 $ & $ 0.324 $ & $ 0.695  $ & $ 0.835 $\\
			\hline
			\multicolumn{6}{c} {\textit { With coordinate convolution }} \\
			\hline			
			Vanilla  & $ 0.699 $ & $ 8.145 $ & $ 0.348 $ & $ 0.587 $ & $ 0.747 $\\
			Explicit DT & $ 0.580 $ & $ 8.566 $ & $ 0.280 $ & $ 0.579 $ & $ 0.807 $\\
			Domain-adapted (ours) &$ \mathbf{0.379} $ & $ \mathbf{7.532} $ & $ \mathbf{0.458} $ & $ \mathbf{0.735}  $ & $ \mathbf{0.856} $\\
			\hline
		\end{tabular}
	}
	\end{center}
\end{table}

\begin{figure}[t]
	\centering
	\includegraphics[width=1\linewidth]{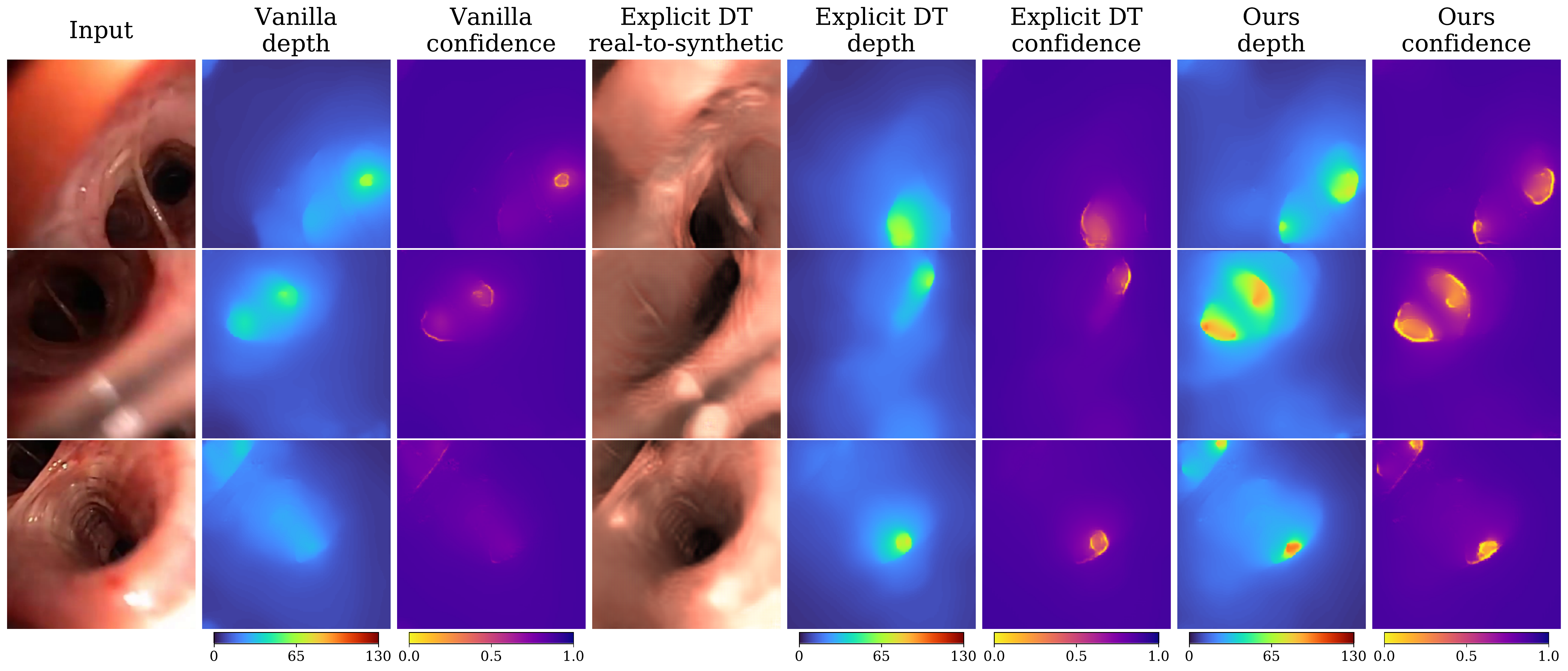}
	\caption{Qualitative analysis on the real images from the human dataset \cite{youtube}. The depth values are in \textit{mm}.}
	\label{fig:eval_human}
\end{figure}

\subsubsection{Qualitative Results.}
In Figure \ref{fig:eval_human}, we display a qualitative comparison of our method against the other approaches on the human pulmonary dataset \cite{youtube}.
Only trained on sharper, rendered synthetic data, we observe that the vanilla network shows difficulties generalizing to the bronchoscope's relatively smoother imaging characteristics.
Additionally, the differences in illumination properties and anatomical discrepancies across the two domains further challenge the vanilla model causing it to fail to assess the deeper regions.
Even though the depth perception of the vanilla model improves using the images that are domain-transferred, the structural inconsistencies generated by CycleGAN become the primary source of error.
Results of our proposed method show that adversarial domain feature adaptation readjusts the encoder to accommodate the characteristics of the real scenes of the human anatomy without the shortcomings of the explicit domain transfer pipelines.
Furthermore, adversarially trained on various patient data, our model shows generalizability to unseen scenes.

During the first-step, supervised training, the vanilla model learns to assign lower confidence to deeper locations and the edges.
We interpret that the first property is caused by the ambiguities in the darker regions, while the latter is supported by virtue of the scale-invariant gradient loss, introduced in Equation \ref{eq:loss_gradient}.
Analyzing confidence estimates in Figure \ref{fig:eval_human}, we see that our model obliges to these characteristics more strictly than the others.
This affirms that our adversarial training step correctly adapts the vanilla model's encoder for more reliable feature extraction at the new domain, ultimately enhancing its depth perception.

\subsubsection{3D Reconstruction Results.}
\begin{figure}[t]
	\centering
	\includegraphics[width=1\linewidth]{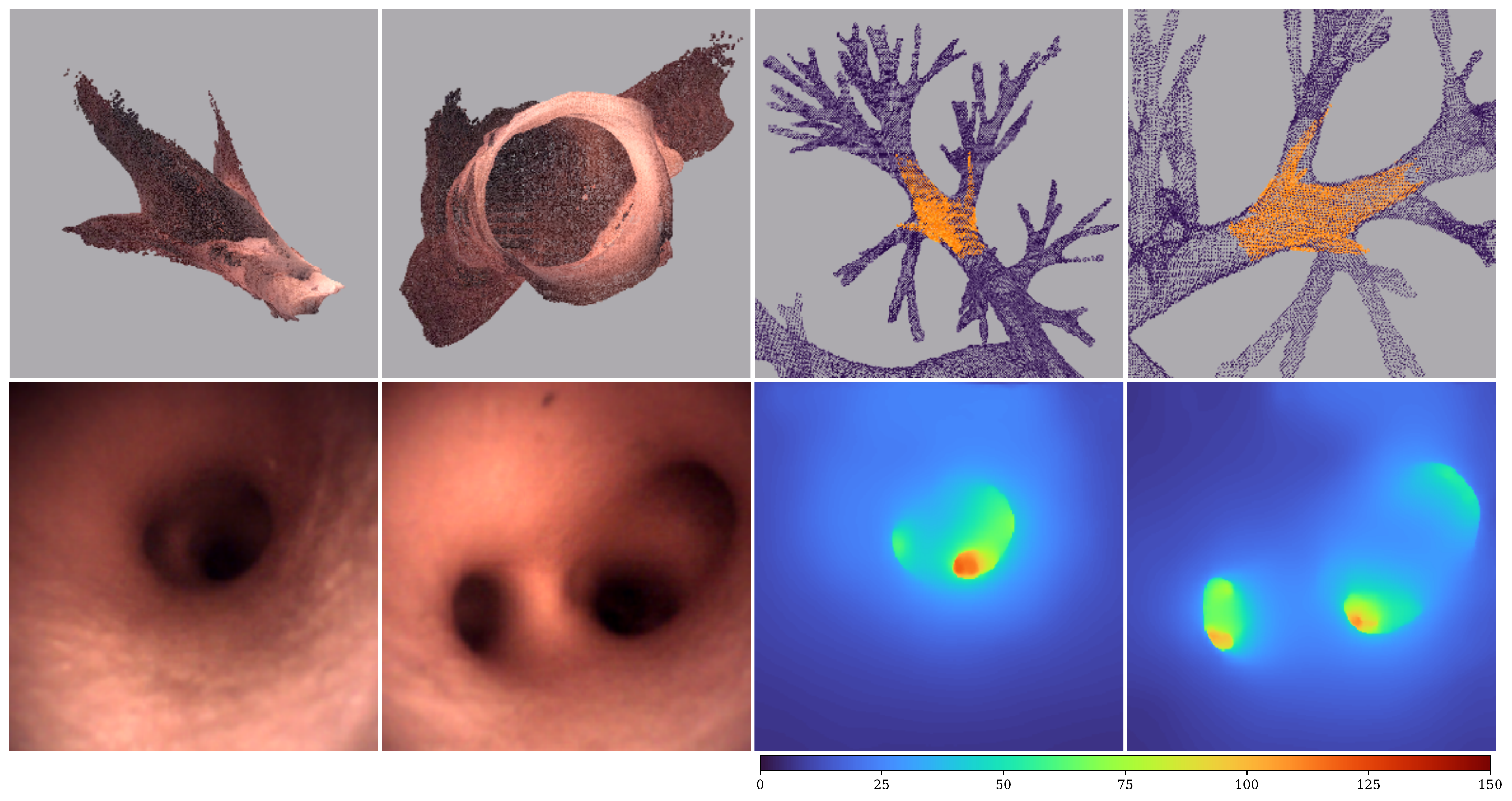}
	\caption{Top-left pair: Point-cloud representation with color information from different perspectives. Top-right pair: Manual scaling and registration of the reconstruction (orange) to the pulmonary volume (purple) from different perspectives. Bottom-left pair: The first and the last frames of the input color sequence. Bottom-right pair: Corresponding depth estimation results using our approach in \textit{mm}. }
	\label{fig:eval_pcloud}
\end{figure}

For this experiment, we use a sequence of 300 frames recorded inside the pulmonary phantom starting from a midpoint of a bronchus and pivot forward towards a bifurcation point. 
We estimate the depth images using our model to create a set of RGB-D frames.
Figure \ref{fig:eval_pcloud} displays the point cloud output of the framework, manually scaled and registered to the point cloud of the phantom extracted from its CT scan.
The results reveal that our method adheres well to the scene's structure, with a certain amount of outliers.
\section{Conclusion}
This paper approaches the problem of depth estimation in bronchoscopic scenes from an adversarial domain feature adaptation perspective.
Our novel pipeline compensates for the lack of labeled bronchoscopy data and the pulmonary system's feature-scarce anatomy by utilizing a two-step training scheme.
The quantitative results show that our method's accuracy is higher than the model that is only trained on the synthetic dataset and the CycleGAN based explicit domain transfer approach we have implemented for comparison.
Our qualitative results demonstrate that the proposed method preserves the scene's structure while the CycleGAN often fails during the real to synthetic transformation.
The experiments reveal that the adopted adversarial training approach is capable of improving over the base model, accomodating for various sources of domain gaps such as illumination, sensory, and anatomical discrepancies.
Furthermore, we show that our approach is capable of correctly reconstructing chunks of bronchoscopic sequences. 

In the future, our primary objective will be to increase the quantitative tests for a more comprehensive evaluation.
Ultimately, we will focus on improving and integrating our method into a SLAM-based navigation pipeline tailored for pulmonary anatomy.

%
%
%
\bibliographystyle{splncs04}
\bibliography{MICCAI21-1715-Bibliography}

\title{Supplementary Material}
\titlerunning{Adversarial Domain Feature Adaptation for Bronchoscopic Depth}
\author{}

\authorrunning{M.A. Karaoglu et al.}
\tocauthor{M.A. Karaoglu}
\institute{}
%
%
\maketitle              
%

\definecolor{Input}{HTML}{FFFFFF}
\definecolor{Conv7}{HTML}{c85da3}
\definecolor{Maxpool3}{HTML}{a5933e}
\definecolor{Resblock}{HTML}{cc5156}
\definecolor{RefpadCoordconvElu}{HTML}{00a1d4}
\definecolor{Refpadconv3Elu}{HTML}{4ea77e}
\definecolor{Refpadconv3Relu}{HTML}{ce783c}
\definecolor{Refpadconv3Sigmoid}{HTML}{846dca}

\tikzset{Input/.style={black,draw=black,fill=Input,rectangle,minimum height=0.1cm}}
\tikzset{Conv7/.style={black,draw=black,fill=Conv7,rectangle,minimum height=0.1cm}}
\tikzset{Maxpool3/.style={black,draw=black,fill=Maxpool3,rectangle,minimum height=0.1cm}}
\tikzset{Resblock/.style={black,draw=black,fill=Resblock,rectangle,minimum height=0.1cm}}
\tikzset{RefpadCoordconvElu/.style={black,draw=black,fill=RefpadCoordconvElu,rectangle,minimum height=0.1cm}}
\tikzset{Refpadconv3Elu/.style={black,draw=black,fill=Refpadconv3Elu,rectangle,minimum height=0.1cm}}
\tikzset{Refpadconv3Relu/.style={black,draw=black,fill=Refpadconv3Relu,rectangle,minimum height=0.1cm}}
\tikzset{Refpadconv3Sigmoid/.style={black,draw=black,fill=Refpadconv3Sigmoid,rectangle,minimum height=0.1cm}}

\section{Details on the Network Architectures}
\begin{figure} []
	\centering
	\begin{tikzpicture}	[thick, scale=0.9, every node/.style={scale=0.9}]
		\node[Input, label={[xshift=0.29cm, yshift=-3.4cm]\tiny 3}, rotate=90,minimum width=6.4cm, minimum height=0.1cm] (input) at (0,0) {};
		\node[Conv7, label={[xshift=0.35cm, yshift=-1.8cm]\tiny 64}, rotate=90,minimum width=3.2cm, minimum height=0.1cm] (encoder0) at (0.4, -1.6) {};
		\node[RefpadCoordconvElu, rotate=90,minimum width=3.2cm, minimum height=0.1cm] (coord0) at (0.8, -1.6) {};
		
		\node[Maxpool3,rotate=90,minimum width=1.6cm, minimum height=0.1cm] (encoder1) at (1.2, -2.4) {};
		
		\node[Resblock, rotate=90,minimum width=1.6cm, minimum height=0.1cm] (encoder2) at (1.6, -2.4) {};
		\node[Resblock, rotate=90,minimum width=1.6cm, minimum height=0.1cm] (encoder3) at (2, -2.4) {};
		\node[RefpadCoordconvElu, rotate=90,minimum width=1.6cm, minimum height=0.1cm] (coord1) at (2.4, -2.4) {};
		
		\node[Resblock,rotate=90, label={[xshift=0.41cm, yshift=-0.6cm]\tiny 128}, minimum width=0.8cm, minimum height=0.1cm] (encoder4) at (2.8, -2.8) {};
		\node[Resblock,rotate=90, minimum width=0.8cm, minimum height=0.1cm] (encoder5) at (3.2, -2.8) {};
		\node[RefpadCoordconvElu,draw=red, rotate=90,minimum width=0.8cm, minimum height=0.1cm] (coord2) at (3.6, -2.8) {};
		
		\node[Resblock,rotate=90, label={[xshift=0.41cm, yshift=-0.4cm]\tiny 256} ,minimum width=0.4cm, minimum height=0.1cm] (encoder6) at (4, -3.0) {};
		\node[Resblock,rotate=90, minimum width=0.4cm, minimum height=0.1cm] (encoder7) at (4.4, -3.0) {};
		\node[RefpadCoordconvElu,  draw=red,rotate=90,minimum width=0.4cm, minimum height=0.1cm] (coord3) at (4.8, -3.0) {};
		
		\node[Resblock, label={[xshift=0.41cm, yshift=-0.3cm]\tiny 512}, rotate=90,minimum width=0.2cm, minimum height=0.1cm] (encoder8) at (5.2, -3.1) {};
		\node[Resblock, rotate=90,minimum width=0.2cm, minimum height=0.1cm] (encoder9) at (5.6, -3.1) {};
		\node[RefpadCoordconvElu, draw=red, rotate=90,minimum width=0.2cm, minimum height=0.1cm] (coord4) at (6, -3.1) {};
		
		
		\node[Refpadconv3Elu, label={[xshift=0.41cm, yshift=-0.3cm]\tiny 256}, rotate=90,minimum width=0.2cm, minimum height=0.1cm] (decoder0) at (6.4, -3.1) {};
		
		\node[Refpadconv3Elu,rotate=90,minimum width=0.4cm, minimum height=0.1cm] (decoder1) at (6.8, -3.0) {};
		\node[Refpadconv3Elu, label={[xshift=0.41cm, yshift=-0.4cm]\tiny 128},rotate=90,minimum width=0.4cm, minimum height=0.1cm] (decoder2) at (7.2, -3.0) {};
		
		\node[Refpadconv3Elu,rotate=90,minimum width=0.8cm, minimum height=0.1cm] (decoder3) at (7.6, -2.8) {};
		\node[Refpadconv3Elu, label={[xshift=0.35cm, yshift=-0.6cm]\tiny 64},rotate=90,minimum width=0.8cm, minimum height=0.1cm] (decoder4) at (8, -2.8) {};
		\node[Refpadconv3Relu, label={[xshift=0.29cm, yshift=-0.6cm]\tiny 1}, rotate=90,minimum width=0.8cm, minimum height=0.1cm] (depth0) at (8.3, -2.8) {};
		\node[Refpadconv3Sigmoid, label={[xshift=0.29cm, yshift=-0.6cm]\tiny 1},rotate=90,minimum width=0.8cm, minimum height=0.1cm] (confidence0) at (8.6, -2.8) {};
		
		\node[Refpadconv3Elu,rotate=90,minimum width=1.6cm, minimum height=0.1cm] (decoder5) at (9, -2.4) {};
		\node[Refpadconv3Elu, label={[xshift=0.35cm, yshift=-1cm]\tiny 32},rotate=90,minimum width=1.6cm, minimum height=0.1cm] (decoder6) at (9.4, -2.4) {};
		\node[Refpadconv3Relu, label={[xshift=0.29cm, yshift=-1cm]\tiny 1},rotate=90,minimum width=1.6cm, minimum height=0.1cm] (depth1) at (9.7, -2.4) {};
		\node[Refpadconv3Sigmoid, label={[xshift=0.29cm, yshift=-1cm]\tiny 1},rotate=90,minimum width=1.6cm, minimum height=0.1cm] (confidence1) at (10, -2.4) {};
		
		\node[Refpadconv3Elu,rotate=90,minimum width=3.2cm, minimum height=0.1cm] (decoder7) at (10.4, -1.6) {};
		\node[Refpadconv3Elu, label={[xshift=0.35cm, yshift=-1.8cm]\tiny 16}, rotate=90,minimum width=3.2cm, minimum height=0.1cm] (decoder8) at (10.8, -1.6) {};
		\node[Refpadconv3Relu, label={[xshift=0.29cm, yshift=-1.8cm]\tiny 1} ,rotate=90,minimum width=3.2cm, minimum height=0.1cm] (depth2) at (11.1, -1.6) {};
		\node[Refpadconv3Sigmoid, label={[xshift=0.29cm, yshift=-1.8cm]\tiny 1}, rotate=90,minimum width=3.2cm, minimum height=0.1cm] (confidence2) at (11.4, -1.6) {};
		
		\node[Refpadconv3Elu,rotate=90,minimum width=6.4cm, minimum height=0.1cm] (decoder9) at (11.8,0) {};
		\node[Refpadconv3Relu, label={[xshift=0.29cm, yshift=-3.4cm]\tiny 1},rotate=90,minimum width=6.4cm, minimum height=0.1cm] (depth3) at (12.2,0) {};
		\node[Refpadconv3Sigmoid, label={[xshift=0.29cm, yshift=-3.4cm]\tiny 1}, rotate=90,minimum width=6.4cm, minimum height=0.1cm] (confidence3) at (12.5,0) {};
		
		
		\coordinate (cinput) at ($ (input) - (-0.1, 1.6) $);
		\draw[black, arrows={-Triangle[scale=0.5]}]  (cinput) -- (encoder0);
		
		\draw[black, arrows={-Triangle[scale=0.5]}]  (encoder0) -- (coord0);
		
		\coordinate (cencoder00) at ($ (encoder0) - (0, 1.6) $);
		\coordinate (cencoder01) at ($ (encoder0) - (0, 1.7) $);
		\coordinate (cencoder10) at ($ (encoder1) - (0, 0.8) $);
		\coordinate (cencoder11) at ($ (encoder1) - (0, 0.9) $);
		\draw[-] (cencoder00) -- (cencoder01);
		\draw[-] (cencoder01) -- (cencoder11);
		\draw[black, arrows={-Triangle[scale=0.5]}]  (cencoder11) -- (cencoder10);
		
		\coordinate (cencoder1) at ($ (encoder1) - (-0.1, 0) $);
		\draw[black, arrows={-Triangle[scale=0.5]}]  (cencoder1) -- (encoder2);
		
		\coordinate (cencoder2) at ($ (encoder2) - (-0.1, 0) $);
		\draw[black, arrows={-Triangle[scale=0.5]}]  (cencoder2) -- (encoder3);
		
		\draw[black, arrows={-Triangle[scale=0.5]}]  (encoder3) -- (coord1);
		
		\coordinate (cencoder30) at ($ (encoder3) - (0, 0.8) $);
		\coordinate (cencoder31) at ($ (encoder3) - (0, 0.9) $);
		\coordinate (cencoder40) at ($ (encoder4) - (0, 0.4) $);
		\coordinate (cencoder41) at ($ (encoder4) - (0, 0.5) $);
		\draw[-] (cencoder30) -- (cencoder31);
		\draw[-] (cencoder31) -- (cencoder41);
		\draw[black, arrows={-Triangle[scale=0.5]}]  (cencoder41) -- (cencoder40);
		
		\coordinate (cencoder4) at ($ (encoder4) - (-0.1, 0) $);
		\draw[black, arrows={-Triangle[scale=0.5]}] (cencoder4) -- (encoder5);
		
		\draw[black, arrows={-Triangle[scale=0.5]}]  (encoder5) -- (coord2);
		
		\coordinate (cencoder50) at ($ (encoder5) - (0, 0.4) $);
		\coordinate (cencoder51) at ($ (encoder5) - (0, 0.5) $);
		\coordinate (cencoder60) at ($ (encoder6) - (0, 0.2) $);
		\coordinate (cencoder61) at ($ (encoder6) - (0, 0.3) $);
		\draw[-] (cencoder50) -- (cencoder51);
		\draw[-] (cencoder51) -- (cencoder61);
		\draw[black, arrows={-Triangle[scale=0.5]}] (cencoder61) -- (cencoder60);
		
		\coordinate (cencoder6) at ($ (encoder6) - (-0.1, 0) $);
		\draw[black, arrows={-Triangle[scale=0.5]}]  (cencoder6) -- (encoder7);
		
		\draw[black, arrows={-Triangle[scale=0.5]}]  (encoder7) -- (coord3);
		
		\coordinate (cencoder70) at ($ (encoder7) - (0, 0.2) $);
		\coordinate (cencoder71) at ($ (encoder7) - (0, 0.3) $);
		\coordinate (cencoder80) at ($ (encoder8) - (0, 0.1) $);
		\coordinate (cencoder81) at ($ (encoder8) - (0, 0.2) $);
		\draw[-] (cencoder70) -- (cencoder71);
		\draw[-] (cencoder71) -- (cencoder81);
		\draw[black, arrows={-Triangle[scale=0.5]}]  (cencoder81) -- (cencoder80);
		
		\coordinate (cencoder8) at ($ (encoder8) - (-0.1, 0) $);
		\draw[black, arrows={-Triangle[scale=0.5]}]  (cencoder8) -- (encoder9);
		
		\draw[black, arrows={-Triangle[scale=0.5]}]  (encoder9) -- (coord4);
		
		\coordinate (ccoord4) at ($ (coord4) - (-0.1, 0) $);
		\draw[black, arrows={-Triangle[scale=0.5]}] (ccoord4) -- (decoder0);
		
		\coordinate (cdecoder2) at ($ (decoder2) - (0.1, 0) $);
		\draw[black, arrows={-Triangle[scale=0.5]}] (decoder1) -- (cdecoder2);
		
		\coordinate (cdecoder4) at ($ (decoder4) - (0.1, 0) $);
		\draw[black, arrows={-Triangle[scale=0.5]}]  (decoder3) -- (cdecoder4);
		
		\coordinate (cdecoder6) at ($ (decoder6) - (0.1, 0) $);
		\draw[black, arrows={-Triangle[scale=0.5]}] (decoder5) -- (cdecoder6);
		
		\coordinate (cdecoder8) at ($ (decoder8) - (0.1, 0) $);
		\draw[black, arrows={-Triangle[scale=0.5]}]  (decoder7) -- (cdecoder8);
		
		\coordinate (cdepth3) at ($ (depth3) - (0.1, 0) $);
		\draw[black, arrows={-Triangle[scale=0.5]}]  (decoder9) -- (cdepth3);
		
		
		\coordinate (cskipcoord3) at ($ (coord3) - (-0.1, -0.1) $);
		\coordinate (cskipdecoder1) at ($ (decoder1) - (0.1, -0.1) $);
		\draw[black, arrows={-Triangle[scale=0.5]}]  (cskipcoord3) -- (cskipdecoder1);
		
		\coordinate (cskipcoord2) at ($ (coord2) - (-0.1, -0.2) $);
		\coordinate (cskipdecoder3) at ($ (decoder3) - (0.1, -0.2) $);
		\draw[black, arrows={-Triangle[scale=0.5]}]  (cskipcoord2) -- (cskipdecoder3);
		
		\coordinate (cskipcoord1) at ($ (coord1) - (-0.1, -0.4) $);
		\coordinate (cskipdecoder5) at ($ (decoder5) - (0.1, -0.4) $);
		\draw[black, arrows={-Triangle[scale=0.5]}] (cskipcoord1) -- (cskipdecoder5);
		
		\coordinate (cskipcoord0) at ($ (coord0) - (-0.1, -0.8) $);
		\coordinate (cskipdecoder7) at ($ (decoder7) - (0.1, -0.8) $);
		\draw[black, arrows={-Triangle[scale=0.5]}] (cskipcoord0) -- (cskipdecoder7);
		
		
		\coordinate (cupdecoder00) at ($ (decoder0) - (0, -0.1) $);
		\coordinate (cupdecoder01) at ($ (decoder0) - (0, -0.2) $);
		\draw[red, arrows={-Triangle[scale=0.5]}] (cupdecoder00) -- (cupdecoder01);
		
		\coordinate (cupdecoder20) at ($ (decoder2) - (0, -0.2) $);
		\coordinate (cupdecoder21) at ($ (decoder2) - (0, -0.4) $);
		\draw[red, arrows={-Triangle[scale=0.5]}] (cupdecoder20) -- (cupdecoder21);
		
		\coordinate (cupdecoder40) at ($ (decoder4) - (0, -0.4) $);
		\coordinate (cupdecoder41) at ($ (decoder4) - (0, -0.8) $);
		\draw[red, arrows={-Triangle[scale=0.5]}] (cupdecoder40) -- (cupdecoder41);
		
		\coordinate (cupdecoder60) at ($ (decoder6) - (0, -0.8) $);
		\coordinate (cupdecoder61) at ($ (decoder6) - (0, -1.6) $);
		\draw[red, arrows={-Triangle[scale=0.5]}] (cupdecoder60) -- (cupdecoder61);
		
		\coordinate (cupdecoder80) at ($ (decoder8) - (0, -1.6) $);
		\coordinate (cupdecoder81) at ($ (decoder8) - (0, -3.2) $);
		\draw[red, arrows={-Triangle[scale=0.5]}] (cupdecoder80) -- (cupdecoder81);
		
		\coordinate (cdecoder9) at ($ (decoder9) - (0.1, -1.6) $);
		\draw[black, arrows={-Triangle[scale=0.5]}] (cupdecoder81) -- (cdecoder9);
		
		
		\matrix [below, nodes={font=\tiny}] at (current bounding box.north) {
			\node [Input,label=right:{Input image}] {}; &
			\node [Conv7,label=right:{Conv2D $K\mathpunct{:}7 \, S\mathpunct{:}2\, P\mathpunct{:}3\text(z)$, BatchNorm2D, ReLU}] {}; \\
			\node [Maxpool3,label=right:{MaxPool2D $K\mathpunct{:}3 \, S\mathpunct{:}2\, P\mathpunct{:}1\text(z)$}] {}; &
			\node [Resblock,label=right:{ResidualBlock}] {}; \\
			\node [RefpadCoordconvElu,label=right:{CoordConv2D $K\mathpunct{:}3 \, S\mathpunct{:}1\, P\mathpunct{:}1\text(r)$, ELU}] {}; &
			\node [Refpadconv3Elu,label=right:{Conv2D $K\mathpunct{:}3 \, S\mathpunct{:}1\, P\mathpunct{:}1\text(r)$, ELU}] {}; \\
			\node [Refpadconv3Relu,label=right:{Conv2D $K\mathpunct{:}3 \, S\mathpunct{:}1\, P\mathpunct{:}1\text(r)$, ReLU}] {}; &
			\node [Refpadconv3Sigmoid,label=right:{Conv2D $K\mathpunct{:}3 \, S\mathpunct{:}1\, P\mathpunct{:}1\text(r)$, Sigmoid}] {}; \\
		};
		
	\end{tikzpicture}
	\caption{An illustration of the encoder-decoder architecture. Red edges represent the nearest neighbor upsampling of the features before concatenation with the skip connection. The numbers below the layers are the channel size of the output of each layer. This value is kept the same for the consecutive layers until it changes. The output of the coordinate convolution layers that are highlighted with red borders are the connecting points of the discriminators for the adversarial training. $K$ is the kernel size, $S$ is the stride, and $P$ is the padding size with zero $\text(z)$ or reflection values $\text(r)$. }
	\label{fig:encoderdecoder}
\end{figure}
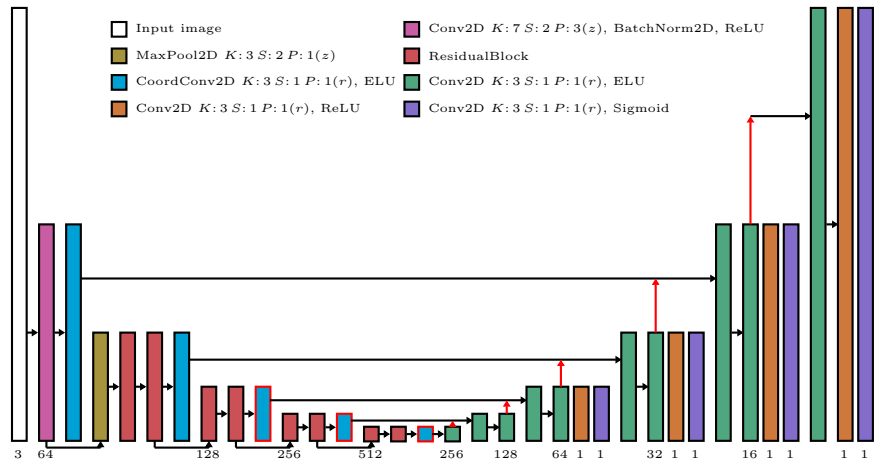

Figure \ref{fig:encoderdecoder} and \ref{fig:discriminator} provide detailed information on the building blocks of our encoder-decoder and discriminator networks.

\definecolor{Input}{HTML}{FFFFFF}
\definecolor{Conv4S2}{HTML}{00a1d4}
\definecolor{Conv3S2}{HTML}{846dca}
\definecolor{LeakyReLU}{HTML}{4ea77e}
\definecolor{InstanceNorm}{HTML}{cc5156}

\tikzset{Input/.style={black,draw=black,fill=Input,rectangle,minimum height=0.1cm}}
\tikzset{Conv4S2/.style={black,draw=black,fill=Conv4S2,rectangle,minimum height=0.1cm}}
\tikzset{Conv3S2/.style={black,draw=black,fill=Conv3S2,rectangle,minimum height=0.1cm}}
\tikzset{LeakyReLU/.style={black,draw=black,fill=LeakyReLU,rectangle,minimum height=0.1cm}}
\tikzset{InstanceNorm/.style={black,draw=black,fill=InstanceNorm,rectangle,minimum height=0.1cm}}

\begin{figure}[]
	\centering
	\begin{tikzpicture}	[thick, scale=1, every node/.style={scale=1}]
		
		
		\node[Input, label={[xshift=0.29cm, yshift=-1.8cm]\tiny n}, rotate=90,minimum width=3.2cm, minimum height=0.1cm] (input) at (0,0) {};
		
		\node[Conv4S2, label={[xshift=0.35cm, yshift=-1.5cm]\tiny 64}, rotate=90,minimum width=2.6cm, minimum height=0.1cm] (conv0) at (0.4, 0) {};
		\node[LeakyReLU, rotate=90,minimum width=2.6cm, minimum height=0.1cm] (leakyrelu0) at (0.8, -0) {};
		
		\node[Conv4S2, label={[xshift=0.41cm, yshift=-1.2cm]\tiny 128}, rotate=90,minimum width=2.0cm, minimum height=0.1cm] (conv1) at (1.2, 0) {};
		\node[InstanceNorm, rotate=90,minimum width=2cm, minimum height=0.1cm] (instancenorm1) at (1.6, -0) {};
		\node[LeakyReLU, rotate=90,minimum width=2cm, minimum height=0.1cm] (leakyrelu1) at (2.0, -0) {};
		
		\node[Conv3S2, label={[xshift=0.41cm, yshift=-0.9cm]\tiny 256}, rotate=90,minimum width=1.4cm, minimum height=0.1cm] (conv2) at (2.4, 0) {};
		\node[InstanceNorm, rotate=90,minimum width=1.4cm, minimum height=0.1cm] (instancenorm2) at (2.8, -0) {};
		\node[LeakyReLU, rotate=90,minimum width=1.4cm, minimum height=0.1cm] (leakyrelu2) at (3.2, -0) {};
		
		\node[Conv3S2, label={[xshift=0.41cm, yshift=-0.6cm]\tiny 512}, rotate=90,minimum width=0.8cm, minimum height=0.1cm] (conv3) at (3.6, 0) {};
		\node[InstanceNorm, rotate=90,minimum width=0.8cm, minimum height=0.1cm] (instancenorm3) at (4.0, -0) {};
		\node[LeakyReLU, rotate=90,minimum width=0.8cm, minimum height=0.1cm] (leakyrelu3) at (4.4, -0) {};
		
		\node[Conv3S2, label={[xshift=0.29cm, yshift=-0.3cm]\tiny 1}, rotate=90,minimum width=0.2cm, minimum height=0.1cm] (conv4) at (4.8, 0) {};
		
		
		\draw[black, arrows={-Triangle[scale=0.5]}]  (input) -- (conv0);
		\draw[black, arrows={-Triangle[scale=0.5]}]  (conv0) -- (leakyrelu0);
		\draw[black, arrows={-Triangle[scale=0.5]}]  (leakyrelu0) -- (conv1);
		\draw[black, arrows={-Triangle[scale=0.5]}]  (conv1) -- (instancenorm1);
		\draw[black, arrows={-Triangle[scale=0.5]}]  (instancenorm1) -- (leakyrelu1);
		\draw[black, arrows={-Triangle[scale=0.5]}]  (leakyrelu1) -- (conv2);
		\draw[black, arrows={-Triangle[scale=0.5]}]  (conv2) -- (instancenorm2);
		\draw[black, arrows={-Triangle[scale=0.5]}]  (instancenorm2) -- (leakyrelu2);
		\draw[black, arrows={-Triangle[scale=0.5]}]  (leakyrelu2) -- (conv3);
		\draw[black, arrows={-Triangle[scale=0.5]}]  (conv3) -- (instancenorm3);
		\draw[black, arrows={-Triangle[scale=0.5]}]  (instancenorm3) -- (leakyrelu3);
		\draw[black, arrows={-Triangle[scale=0.5]}]  (leakyrelu3) -- (conv4);
		
		
		\matrix [below, nodes={font=\tiny}] at (7, 1.8) {
			\node [Input,label=right:{Input features}] {}; \\
			\node [Conv4S2,label=right:{Conv2D $K\mathpunct{:}4 \, S\mathpunct{:}2\, P\mathpunct{:}1\text(z)$}] {}; \\
			\node [Conv3S2,label=right:{Conv2D $K\mathpunct{:}3 \, S\mathpunct{:}1\, P\mathpunct{:}1\text(z)$}] {}; \\
			\node [LeakyReLU,label=right:{LeakyReLU}] {}; \\
			\node [InstanceNorm,label=right:{InstanceNorm2D}] {}; \\
		};	
		
	\end{tikzpicture}
	\caption{An illustration of the discriminator architecture. The numbers below the layers are the channel size of the output of each layer. This value is kept the same for the consecutive layers until it changes. $n$ is the channel size of the input features, $K$ is the kernel size, $S$ is the stride, and $P$ is the padding size with zero values $\text(z)$.}
	\label{fig:discriminator}
\end{figure}
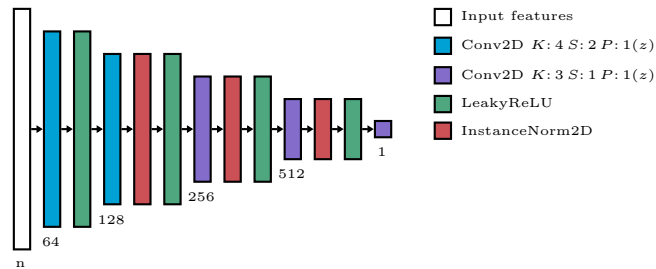

\end{document}